\documentclass[prd,twocolumn,showpacs,preprintnumbers,nofootinbib]{revtex4}
\usepackage[dvipdfmx]{graphicx}
\usepackage{enumerate}
\usepackage{amsmath,amssymb}
\usepackage{mathrsfs}
\usepackage{bm}

\begin{document}

\preprint{CHIBA-EP-219v4, 2016.09.21}
\title{
Gauge-invariant description of Higgs phenomenon and   quark confinement

}

\author{Kei-Ichi Kondo} 
\email{kondok@faculty.chiba-u.jp}
\affiliation{Department of Physics,  Faculty of Science, Chiba University, Chiba 263-8522, Japan}


\begin{abstract}
We propose a novel description for the Higgs mechanism by which a gauge boson acquires the mass.
We do not assume spontaneous breakdown of gauge symmetry signaled by a non-vanishing vacuum expectation value of the scalar field. 
In fact, we give a manifestly gauge-invariant description of the Higgs mechanism in the operator level, which does not rely on spontaneous symmetry breaking. 
This enables us to discuss the  confinement-Higgs complementarity from a new perspective.
The ``Abelian'' dominance in quark confinement of the Yang-Mills theory is understood as a consequence of the gauge-invariant Higgs phenomenon for the relevant  Yang-Mills-Higgs model.
\end{abstract}


\maketitle


\section{Introduction}
\setcounter{equation}{0}

The \textit{Brout-Englert-Higgs  mechanism} or Higgs phenomenon for short is one of the most well-known mechanisms by which gauge bosons \cite{YM54} acquire their masses \cite{Higgs1,Higgs2,Higgs3}. 
In the conventional wisdom, the Higgs mechanism is understood in such a way that the \textit{spontaneous symmetry breaking} (SSB) generates mass for a gauge boson: 
The original gauge group $G$ is spontaneously broken down to a subgroup $H$ by choosing a specific vacuum  as the physical state from all the possible degenerate ground states (the lowest energy states).  
Such  SSB of the original gauge symmetry is caused by a non-vanishing vacuum expectation value (VEV) $\langle \bm{\phi} \rangle \ne 0$ of a scalar field $\bm{\phi}$ governed by a given potential $V(\bm{\phi})$. 
For a continuous group $G$, there appear the massless \textit{Nambu-Goldstone bosons} associated with the SSB $G \to H$ according to the Nambu-Goldstone theorem \cite{Nambu,Goldstone}. 
When the scalar field couples to a gauge field, however,
the massless Nambu-Goldstone bosons are absorbed to provide the gauge boson with the mass. 
Thus, the massless Nambu-Goldstone bosons disappear from the spectrum. 
In a semi-classical treatment, the VEV $\langle \bm{\phi} \rangle$ is identified with one of the minima $\bm{\phi}_0$ of the scalar potential $V(\bm{\phi})$, namely, $\langle \bm{\phi} \rangle=\bm{\phi}_0 \ne 0$ with $V^\prime(\bm{\phi}_0)=0$.

Although this paper focuses on the Higgs phenomenon in the continuum space time, it is very instructive to learn the lattice results, because some non-perturbative and rigorous results are available on the lattice.  Especially, the lattice gauge theory $\grave{a}\  la$ Wilson \cite{Wilson74} gives  a well-defined gauge theory without gauge fixing. 
The \textit{Elitzur theorem} \cite{Elitzur75} tells us that the  local continuous gauge symmetry cannot break spontaneously, if no gauge fixing  is introduced. 
In the absence of gauge fixing, 
all gauge non-invariant Green functions vanish identically. Especially, 
the VEV $\langle \bm{\phi} \rangle$ of the scalar field  $\bm{\phi}$ is rigorously zero, 
\begin{align}
 \langle \bm{\phi} \rangle = 0 ,
\end{align}
no matter what the form of the scalar potential $V(\bm{\phi})$.

Therefore, we are forced to fix the gauge to cause the non-zero VEV. 
Even after the gauge fixing, however, we still have the problem. 
Whether SSB occurs or not depends on the gauge choice. 
For instance, in non-compact $U(1)$ gauge-Higgs model under the covariant gauge fixing with a gauge fixing parameter $\alpha$, the SSB occurs $\langle \bm{\phi} \rangle \ne 0$ only in the Landau gauge $\alpha=0$, and no SSB occur $\langle \bm{\phi} \rangle = 0$ in all other covariant gauges with $\alpha \ne 0$, as rigorously shown in \cite{KK85,BN86}. 
In an axial gauge, $\langle \bm{\phi} \rangle = 0$ for compact models \cite{FMS80}.
In contrast, it can happen that $\langle \bm{\phi} \rangle \ne 0$ in a unitary gauge regardless of the shape of the scalar potential.
It is obvious that the VEV of the scalar field is not a gauge-independent criterion of SSB.

Even after breaking completely the local gauge symmetry $G$ by imposing a suitable gauge fixing condition, there can remain a global gauge symmetry  $H^\prime$ of $G$.
Such a global symmetry $H^\prime$ is called the \textit{remnant global gauge symmetry} \cite{GOZ04,CG08}. 
Only a remnant global gauge symmetry $H^\prime$ of the local gauge symmetry $G$ can break spontaneously to cause the Higgs phenomenon \cite{Kugo81}.  
However, such a subgroup $H^\prime$ is not unique  and the location of the breaking in the phase diagram depends on $H^\prime$ in the gauge-Higgs model. 
The relevant numerical evidences  are given on a lattice \cite{CG08} for different $H^\prime$ allowed for various confinement scenarios.
Moreover, the transition occurs in the regions where the Fradkin-Shenker-Osterwalder-Seiler theorem \cite{FS79,OS78} assures us that there is no transition  in the phase diagram. 
Thus, the spontaneous gauge symmetry breaking is a rather misleading terminology.

These observations indicate that the Higgs phenomenon should be characterized in a gauge-invariant way without breaking the original gauge symmetry.
In this paper, we show that a gauge boson can acquire the mass in a gauge-invariant way without assuming spontaneous breakdown of gauge symmetry which is signaled by the non-vanishing VEV of the scalar field.  
We demonstrate that the Higgs phenomenon occurs even without such SSB. 
The spontaneous symmetry breaking is sufficient but not necessary for the Higgs mechanism to work. 
Remember that quark confinement is realized in the unbroken gauge symmetry phase with \textit{mass gap}. 
Thus, the gauge-invariant description of the Higgs mechanism can shed new light on the \textit{complementarity} between confinement phase and Higgs phase \cite{tHooft80}.


\section{Yang-Mills-Higgs model and the conventional Higgs mechanism}

In this paper we use the notation for the inner product of the Lie-algebra valued quantities $\mathscr{A}=\mathscr{A}^AT_A$ and $\mathscr{B}=\mathscr{B}^AT_A$; $\mathscr{A} \cdot \mathscr{B}:=2{\rm tr}(\mathscr{A}\mathscr{B}) = \mathscr{A}^A \mathscr{B}^B 2{\rm tr}(T_A T_B) = \mathscr{A}^A \mathscr{B}^A$ 
under the normalization ${\rm tr}(T_A T_B) = \frac12 \delta_{AB}$ for the generators $T_A$ of the Lie algebra $su(N)$ ($A=1,2,...,{\rm dim}G=N^2-1$) for a gauge group $G=SU(N)$.
The $SU(N)$ Yang-Mills field $\mathscr{A}_{\mu}(x)=\mathscr{A}_{\mu}^{A}(x)T_A$ has the field strength 
$\mathscr{F}_{\mu\nu}(x)=\mathscr{F}_{\mu\nu}^{A}(x)T_A$  defined by
$
\mathscr{F}_{\mu\nu}  := \partial_{\mu} \mathscr{A}_{\nu} - \partial_{\nu} \mathscr{A}_{\mu} -i g [\mathscr{A}_{\mu}, \mathscr{A}_{\nu}] 
$.

We consider a Yang-Mills-Higgs theory specified by a  gauge-invariant action.
The Yang-Mills field $\mathscr{A}_\mu(x)=\mathscr{A}_\mu^A(x)T_A$ and the \textit{adjoint scalar field} $\bm{\phi}(x)=\phi^A(x)T_A$ obey the gauge transformation:
\begin{align}
    \mathscr{A}_\mu(x) &\to U(x) \mathscr{A}_\mu(x) U^{-1}(x) + ig^{-1} U(x) \partial_\mu U^{-1}(x) ,
\nonumber\\ 
  \bm{\phi}(x) &\to U(x) \bm{\phi}(x) U^{-1}(x) ,
\quad U(x) \in G=SU(N)  .
  \label{gauge-transf}
\end{align}
For concreteness, consider the $G=SU(N)$ Yang-Mills-Higgs theory  with the Lagrangian density:
\begin{align}
\mathscr{L}_{\rm YMH} =& - \frac{1}{4} \mathscr{F}^{\mu\nu }(x) \cdot \mathscr{F}_{\mu\nu}(x) 
\nonumber\\&
+\frac{1}{2 } ( \mathscr{D}^{\mu}[\mathscr{A}] \bm{\phi}(x) ) \cdot (\mathscr{D}_{\mu}[\mathscr{A}] \bm{\phi}(x) )
\nonumber\\&
- V(\bm{\phi}(x) \cdot \bm{\phi}(x)) . 
\label{SU2-YMH-1}
\end{align}
where we have defined the covariant derivative 
$
\mathscr{D}_{\mu}[\mathscr{A}]   := \partial_{\mu}  - ig[ \mathscr{A}_{\mu},  \cdot ] 
$
in the adjoint representation.
We assume that the  {adjoint scalar field} $\bm{\phi}(x)=\phi^A(x)T_A$ has the fixed radial length, which is represented by   a constraint:%
\footnote{
After imposing the constraint (\ref{SU2-YMH-1-constraint}), the subsequent argument should hold irrespective of the form of the potential $V$. 
The vacuum manifold in the target space of the scalar field is determined by the minima of the potential $V$,  which also satisfies the constraint (\ref{SU2-YMH-1-constraint}). 
However, there are some options as to when and how the  constraint is incorporated, see e.g., (\ref{mYM3}). The potential is omitted in what follows when any confusion does not occur. 
Moreover, this model is perturbatively non-renormalizable and the non-perturbative treatment is required. 
}
\begin{align}
 \bm{\phi}(x) \cdot \bm{\phi}(x) \equiv \bm{\phi}^A(x)   \bm{\phi}^A(x) = v^2 .
\label{SU2-YMH-1-constraint}
\end{align}
Notice that $\bm{\phi}(x) \cdot \bm{\phi}(x)$ is a gauge-invariant combination.
Therefore, the potential $V$ as an arbitrary function of $\bm{\phi}(x) \cdot \bm{\phi}(x)$  is invariant under the gauge transformation. 
The covariant derivative 
$
\mathscr{D}_{\mu}[\mathscr{A}]   := \partial_{\mu}  - ig[ \mathscr{A}_{\mu},  \cdot ] 
$
transforms according to the adjoint representation under the gauge transformation: 
$\mathscr{D}_{\mu}[\mathscr{A}]  \to U(x) \mathscr{D}_{\mu}[\mathscr{A}]  U^{-1}(x)$. 
This is also the case for the field strength $\mathscr{F}_{\mu\nu}(x) $. 
Moreover, the constraint (\ref{SU2-YMH-1-constraint}) is invariant under the  gauge transformation and does not break the gauge invariance of the theory.  
Therefore,  $\mathscr{L}_{\rm YMH}$ of (\ref{SU2-YMH-1}) with the constraint (\ref{SU2-YMH-1-constraint}) is  invariant under the local gauge transformation (\ref{gauge-transf}).

For $N=2$, this theory is nothing but the well-known Georgi-Glashow model which exemplifies the SSB of the local gauge symmetry from $SU(2)$ down to $U(1)$ except for the magnitude of the scalar field being fixed (\ref{SU2-YMH-1-constraint}). 
In this paper,  we focus our discussions on the $SU(2)$ case.

First, we recall the conventional description  
for the Higgs mechanism.  
If the scalar field $\bm{\phi}(x)$ acquires a non-vanishing VEV $\langle \bm{\phi}(x) \rangle=\langle \bm{\phi}  \rangle$, then  the covariant derivative reduces to 
\begin{align}
  \mathscr{D}_{\mu}[\mathscr{A}]\phi (x)  
  :=& \partial_{\mu} \bm{\phi}(x) - ig[ \mathscr{A}_{\mu}(x),  \bm{\phi}(x) ] 
  \nonumber\\  
 \to &
- ig [   \mathscr{A}_{\mu}(x),  \langle \bm{\phi} \rangle ]  + ...,
\end{align}
 and the Lagrangian density reads  
\begin{align}
\mathscr{L}_{\rm YMH} \to 
& - \frac{1}{2} {\rm tr}_{G} \{ \mathscr{F}^{\mu\nu }(x) \mathscr{F}_{\mu\nu}(x) \} 
\nonumber\\&
- g^2 {\rm tr}_{G} \{  [\mathscr{A}^{\mu}(x), \langle \bm{\phi} \rangle ]  [\mathscr{A}_{\mu}(x), \langle \bm{\phi} \rangle ] \}    + ... . 
\nonumber\\ 
=& - \frac{1}{2} {\rm tr}_{G} \{ \mathscr{F}^{\mu\nu }(x) \mathscr{F}_{\mu\nu}(x) \} 
\nonumber\\&
- g^2 {\rm tr}_{G} \{  [T_A, \langle \bm{\phi} \rangle ]  [T_B, \langle \bm{\phi} \rangle ] \}  \mathscr{A}^{\mu A}(x)  \mathscr{A}_{\mu }^B(x)  + ... . 
\label{VEV-L}
\end{align}
To break spontaneously  the local continuous gauge symmetry $G$ by realizing the non-vanishing VEV $\langle \bm{\phi} \rangle$ of the scalar field $\bm\phi$, we choose  the \textit{unitary gauge} in which  the scalar field $\bm{\phi}(x)$ is pointed to a specific direction $\bm{\phi}(x) \to \bm{\phi}_{\infty}$ uniformly over the spacetime. 

This procedure does not completely  break the original gauge symmetry $G$. Indeed, there may exist a subgroup $H$ of $G$ such that  $\bm{\phi}_{\infty}$ does not change under the  local $H$ gauge transformation.
This is the \textit{partial SSB} $G \to H$:  
the mass is provided  for the coset $G/H$ (broken parts), while the mass  is not supplied for the $H$-commutative part of   $\mathscr{A}_{\mu}$: 
\begin{align}
\mathscr{L}_{\rm YMH}  \to &  - \frac{1}{2} {\rm tr}_{G} \{ \mathscr{F}^{\mu\nu }(x) \mathscr{F}_{\mu\nu}(x) \}
\nonumber\\&
- (gv)^2 {\rm tr}_{G/H} \{  \mathscr{A}^{\mu}(x)  \mathscr{A}_{\mu}(x)  \} . 
\label{massG/H}
\end{align}
After the partial SSB, therefore, the resulting theory is a gauge theory with the residual gauge group $H$.

For $G=SU(2)$, by taking the usual \textit{unitary gauge} in which the scalar field $\bm{\phi}(x)=\phi^A(x)T_A \ (A=1,2,3)$ is chosen so that  
\begin{align}
\langle \bm{\phi}_{\infty} \rangle = v T_3, 
 \quad {\rm or} \quad 
\langle {\phi}^A_{\infty} \rangle  = v \delta^{A3} ,
   \label{unitary-gauge}
\end{align}
the second term of (\ref{VEV-L}) generates the mass term,
\begin{align}
& - g^2 v^2 {\rm tr}_{G} \{  [T_A, T_3 ]  [T_B, T_3 ] \} \mathscr{A}^{\mu A}(x) \mathscr{A}_{\mu}^B(x) 
\nonumber\\&
=   
\frac{1}{2} g^2 v^2(\mathscr{A}^{\mu 1}(x) \mathscr{A}_{\mu}^{1}(x) + \mathscr{A}^{\mu 2}(x) \mathscr{A}_{\mu}^{2}(x) )
  . 
\end{align}
For $SU(2)$, indeed,  the off-diagonal gluons $\mathscr{A}_{\mu}^{a}$ $(a=1,2)$ 
acquire the same mass $M_{W} :=gv$, while the diagonal gluon $\mathscr{A}_{\mu}^{3}$ remains massless. 
Even after taking the unitary gauge (\ref{unitary-gauge}), $U(1)$ gauge symmetry described by $\mathscr{A}_{\mu}^{3}$ still remains as the residual local gauge symmetry $H=U(1)$, which leaves  $\bm{\phi}_{\infty}$   invariant (the local rotation around the axis  of the scalar field $\bm{\phi}_{\infty}$).

Thus, the SSB is sufficient for the Higgs mechanism to take place. 
But, it is not clear whether the SSB is necessary or not for the Higgs mechanism to work.

In the complete SSB $G \to H=\{ 1 \}$, all components of the Yang-Mills field become massive with no massless components: 
\begin{align}
\mathscr{L}_{\rm YMH} \to & - \frac{1}{2} {\rm tr}_{G} \{ \mathscr{F}^{\mu\nu }  \mathscr{F}_{\mu\nu}  \}
-  (gv)^2  {\rm tr}_{G} \{  \mathscr{A}^{\mu}  \mathscr{A}_{\mu}  \} , 
\end{align}
and the resulting theory has no residual gauge symmetry.
This case should be  separately discussed, see Appendix. 

\section{Gauge-invariant Higgs mechanism: SU(2) case}

Next, we give a novel description, namely, a gauge-invariant (gauge-independent) description of mass generation for gauge bosons without relying on the SSB. 
We construct a composite vector field $\mathscr{W}_\mu(x)$ from the Yang-Mills field $\mathscr{A}_\mu(x)$ and the (adjoint) scalar field $\bm{\phi}(x)$ by 
\begin{align}
 \mathscr{W}_\mu(x) 
:=   -ig^{-1}  
  [\hat{\bm{\phi}}(x), \mathscr{D}_\mu[\mathscr{A}]\hat{\bm{\phi}}(x) ]
 ,
\label{W-def1}
\end{align}
with  the unit scalar field $\hat{\bm{\phi}}$   defined by 
\begin{align}
\hat{\bm{\phi}}(x):=\bm{\phi}(x)/v ,
\end{align}
and  the covariant derivative in the adjoint representation
$
\mathscr{D}_{\mu}[\mathscr{A}]\bm{\phi}  := \partial_{\mu} \bm{\phi} - ig[ \mathscr{A}_{\mu}   \bm{\phi} ] 
$. 
We find that the kinetic term of the Yang-Mills-Higgs model is identical to the ``mass term'' of the vector field $\mathscr{W}_\mu(x)$:
\begin{align}
 \frac{1}{2 }  \mathscr{D}^{\mu}[\mathscr{A}] \bm{\phi}(x) \cdot \mathscr{D}_{\mu}[\mathscr{A}] \bm{\phi}(x)  
=&  \frac{1}{2 } M_{W}^2 \mathscr{W}^\mu(x) \cdot \mathscr{W}_\mu(x) , 
  \nonumber\\
 M_{W} :=& gv 
,
\label{W-mass}
\end{align}
as far as  the constraint (\ref{SU2-YMH-1-constraint})   is satisfied. 
Indeed, this fact is shown explicitly for $G=SU(2)$: 
\begin{align}
  g^2 v^2 \mathscr{W}^\mu \cdot \mathscr{W}_\mu 
=&  v^{-2} 2{\rm tr}( [\bm{\phi} , \mathscr{D}^\mu[\mathscr{A}]\bm{\phi}  ] [\bm{\phi} , \mathscr{D}_\mu[\mathscr{A}]\bm{\phi}  ] )
  \nonumber\\
=& v^{-2} \{ (\bm{\phi} \cdot \bm{\phi}  )  (\mathscr{D}^{\mu}[\mathscr{A}] \bm{\phi} \cdot \mathscr{D}_{\mu}[\mathscr{A}] \bm{\phi}) 
\nonumber\\ &
 - (\bm{\phi} \cdot \mathscr{D}^{\mu}[\mathscr{A}] \bm{\phi})(\bm{\phi} \cdot \mathscr{D}_{\mu}[\mathscr{A}] \bm{\phi}) \}
\nonumber\\ 
=& (\mathscr{D}^{\mu}[\mathscr{A}] \bm{\phi}) \cdot (\mathscr{D}_{\mu}[\mathscr{A}] \bm{\phi}) 
  ,
\end{align}
where we have used the constraint (\ref{SU2-YMH-1-constraint}) and
$
  \bm{\phi} \cdot \mathscr{D}_{\mu}[\mathscr{A}] \bm{\phi}
 =  \bm{\phi} \cdot \partial_\mu \bm{\phi} + \bm{\phi} \cdot (g \mathscr{A}_{\mu} \times \bm{\phi} )   
=    g\mathscr{A}_{\mu} \cdot (\bm{\phi} \times \bm{\phi} )   
 = 0
  ,
$
with $\bm{\phi} \cdot \partial_\mu \bm{\phi}=0$ following from differentiating the constraint (\ref{SU2-YMH-1-constraint}). 

Remarkably, the above ``mass term'' (\ref{W-mass}) of $\mathscr{W}_\mu$ is gauge invariant, since   $\mathscr{W}_\mu$ obeys the adjoint gauge transformation:
\begin{align}
  \mathscr{W}_\mu(x) \to U(x) \mathscr{W}_\mu(x) U^{-1}(x)
  .
\end{align}
Therefore, the vector field $\mathscr{W}_\mu$ becomes massive without breaking the original gauge symmetry. 
The above description shows that the SSB of gauge symmetry is not necessary for generating the mass of gauge bosons $\mathscr{W}_\mu$, since we do not need to choose a specific vacuum from all  possible degenerate ground states distinguished by the direction of $\bm{\phi}$. 
The relation (\ref{W-def1}) gives a gauge-independent definition of the massive gluon mode  in the operator level. 
The relation (\ref{W-def1}) is also \textit{independent from the parameterization of the scalar field}.  See Appendix in which the statement is exemplified for a simpler model.

How is this description related to the conventional one?
The constraint $\bm{\phi} \cdot \bm{\phi} =v^2$ represents the vacuum manifold in the target space of the scalar field $\bm{\phi}$.
The scalar field $\bm{\phi}$ subject to the constraint $\bm{\phi} \cdot \bm{\phi} =v^2$ is regarded as the \textit{Nambu-Goldstone modes} living in the \textit{flat direction} at the bottom of the potential $V(\phi)$, giving the \textit{degenerate lowest energy states}.
Therefore, the massive field $\mathscr{W}_\mu$ is   formed by combining the \textit{massless (would-be) Nambu-Goldstone modes} with the original massless Yang-Mills field $\mathscr{A}_\mu$. 
This corresponds to the  conventional  explanation that the gauge boson acquires the mass by absorbing the Nambu-Goldstone boson appeared in association with the SSB.  

Despite its appearance (\ref{W-def1}) of $\mathscr{W}_\mu$ obeying the adjoint gauge transformation, the independent internal degrees of freedom of the new field $\mathscr{W}_\mu=(\mathscr{W}_\mu^{A})$ $(A=1,2,3)$ is equal to  ${\rm dim}(G/H)=2$, since $\mathscr{W}_\mu$  has no components parallel to the scalar field, that is to say, $\mathscr{W}_\mu$ is orthogonal to the scalar field $\bm{\phi}$:
\begin{align}
 \mathscr{W}_\mu(x) \cdot  \bm{\phi}(x) 
= 0 
 .
\label{W-def2}
\end{align}
Notice that this is a gauge-invariant statement. 
Thus, $\mathscr{W}_\mu(x)$ represent the massive modes corresponding to the coset space $G/H$ as expected. 
In this way, we can understand the \textit{residual gauge symmetry}  left in  the partial SSB: $G=SU(2) \to H=U(1)$. 
In fact, by taking the unitary gauge $\bm{\phi}(x) \to \bm{\phi}_{\infty}=v \hat{\bm{\phi}}_{\infty}$,   $\mathscr{W}_\mu$ reduces to 
\begin{align}
 \mathscr{W}_\mu(x) 
\to & -ig^{-1}  
  [\hat{\bm{\phi}}_{\infty}, \mathscr{D}_\mu[\mathscr{A}]\hat{\bm{\phi}}_{\infty} ] 
\nonumber\\ 
=&   
  [\hat{\bm{\phi}}_{\infty},   [  \hat{\bm{\phi}}_{\infty},  \mathscr{A}_\mu(x)  ]] 
\nonumber\\ 
=& \mathscr{A}_\mu(x) - (\mathscr{A}_\mu(x) \cdot \hat{\bm{\phi}}_{\infty})\hat{\bm{\phi}}_{\infty} 
 .
\end{align}
Then $\mathscr{W}_\mu$ agrees with the off-diagonal components for the specific choice $\hat{\bm{\phi}}_{\infty}^A=\delta^{A3}$:
\begin{align}
 \mathscr{W}_\mu^{A}(x) 
\to   \begin{cases}
 \mathscr{A}_\mu^{a}(x)  &(A=a=1,2) \\
 0 &(A=3)
\end{cases}  .
\end{align}

This suggests that the original gauge field $\mathscr{A}_\mu$ is separated  into two pieces:
\begin{align}
 \mathscr{A}_\mu(x) = \mathscr{V}_\mu(x) + \mathscr{W}_\mu(x)   
 .
 \label{A-decomposition}
\end{align}
By definition, the field $\mathscr{V}_\mu(x)$ transforms  under the  gauge transformation just like the original gauge field $\mathscr{A}_\mu(x)$: 
\begin{align}
  \mathscr{V}_\mu(x) \to U(x) \mathscr{V}_\mu(x) U^{-1}(x) + ig^{-1} U(x) \partial_\mu U^{-1}(x) 
  .
\end{align}
Then the question is how to characterize the first piece $\mathscr{V}_\mu(x)$ which is expected to become dominant in the low-energy $E \ll M_{W}$  region, where $\mathscr{W}_\mu(x)$ with the mass $M_{W}$  can be  negligible.
According to (\ref{W-def1}), it is shown that $\mathscr{W}_\mu(x) = 0$ is equivalent to 
\begin{align}
 \mathscr{D}_\mu[\mathscr{V}] \hat{\bm{\phi}}(x) = 0 
 .
\label{V-def2}
\end{align}
Using the first equation (\ref{W-def2}) and the second equation (\ref{V-def2}), 
we find that a composite vector field $\mathscr{V}_\mu$ is constructed from the Yang-Mills field $\mathscr{A}_\mu$ and the scalar field $\bm{\phi}$ as \cite{Manton74}:
\begin{align}
 \mathscr{V}_\mu(x) =& c_\mu(x)  \hat{\bm{\phi}}(x) + ig^{-1} [  \hat{\bm{\phi}}(x), \partial_\mu \hat{\bm{\phi}}(x)  ]   , 
\label{V-def1}
\\
  & c_\mu(x) :=  \mathscr{A}_\mu(x) \cdot \hat{\bm{\phi}}(x)
 .
\label{c-def1}
\end{align}
In fact, this form for  $\mathscr{V}_\mu(x)$ agrees with $\mathscr{V}_\mu(x)=\mathscr{A}_\mu(x)-\mathscr{W}_\mu(x)$ when eq.(\ref{W-def1}) is substituted into    $\mathscr{W}_\mu(x)$.
In the unitary gauge $\bm{\phi}(x) \to \bm{\phi}_{\infty}=v \hat{\bm{\phi}}_{\infty}$,  $\mathscr{V}_\mu$ reduces to
\begin{align}
 \mathscr{V}_\mu(x)  
\to   ( \mathscr{A}_\mu(x) \cdot \hat{\bm{\phi}}_{\infty} ) \hat{\bm{\phi}}_{\infty} .
\end{align}
Then, $\mathscr{V}_\mu$ agrees with the diagonal component for   $\hat{\bm{\phi}}_{\infty}^A=\delta^{A3}$:
\begin{align}
 \mathscr{V}_\mu^{A}(x) 
\to    \begin{cases}
 0  &(A=a=1,2) \\
 \mathscr{A}_\mu^{3}(x) &(A=3)
\end{cases}  .
\end{align}
Thus, the above arguments go well in the topologically trivial sector.

In the topologically non-trivial sector, the above argument must be improved, since $\partial_\mu \hat{\bm{\phi}}$ is not identically zero in the presence of singularities related to the topological configuration.   
Indeed, in order to realize the unitary gauge configuration starting from the hedgehog configuration of the scalar field, we need to perform the singular gauge transformation  in the presence of the \textit{'t Hooft-Polyakov magnetic monopole} \cite{tHP74}. 
This case will be refined later. 

Notice that the decomposition equality (\ref{A-decomposition}) represents a rather non-trivial statement where  $\mathscr{V}_\mu(x)$ is identified with (\ref{V-def1}) and $\mathscr{W}_\mu(x)$ is identified with (\ref{W-def1}).
We first introduce the fields $\mathscr{V}_\mu(x)$ and $\mathscr{W}_\mu(x)$ as composite field operators of $\mathscr{A}_\mu(x)$ and $\hat{\bm{\phi}}(x)$. 
Then we regard a set of field variables  $\{ c_\mu(x), \mathscr{W}_\mu(x), \hat{\bm{\phi}}(x) \}$ as obtained from  $\{ \mathscr{A}_\mu(x), \hat{\bm{\phi}}(x) \}$ based on  \textit{change of variables}: 
\begin{align}
   \{ c_\mu(x), \mathscr{W}_\mu(x), \hat{\bm{\phi}}(x) \} \leftarrow \{ \mathscr{A}_\mu(x), \hat{\bm{\phi}}(x) \}  ,
\end{align}
where (\ref{c-def1}) and (\ref{W-def1}) give respectively the transformation law of $c_\mu(x)$ and $\mathscr{W}_\mu(x)$  from $\{ \mathscr{A}_\mu(x), \hat{\bm{\phi}}(x) \}$. 
Indeed, we can calculate the Jacobian associated with this change of variables. See  \cite{KMS06,Kondo06,KKSS15} for details. 
Finally, we identify $c_\mu(x)$, $\mathscr{W}_\mu(x)$ and $\hat{\bm{\phi}}(x)$ with the fundamental field variables (which are independent up to the constraint (\ref{SU2-YMH-1-constraint})) for describing the massive  Yang-Mills theory anew. (Here fundamental means that the quantization should be performed with respect to those variables $\{ c_\mu(x), \mathscr{W}_\mu(x), \hat{\bm{\phi}}(x) \}$ which appear e.g., in the path-integral measure.) 

According to the decomposition (\ref{A-decomposition}), the field strength $\mathscr{F}_{\mu\nu}(x)$ of the gauge field $\mathscr{A}_\mu(x)$ is decomposed as  
\begin{align}
  \mathscr{F}_{\mu\nu}[\mathscr{A}] :=& \partial_\mu \mathscr{A}_\nu - \partial_\nu \mathscr{A}_\mu -i g [ \mathscr{A}_\mu , \mathscr{A}_\nu ]
\nonumber\\
=& \mathscr{F}_{\mu\nu}[\mathscr{V}]  + \mathscr{D}_\mu[\mathscr{V}] \mathscr{W}_\nu - \mathscr{D}_\nu[\mathscr{V}] \mathscr{W}_\mu 
\nonumber\\&
-i g [ \mathscr{W}_\mu , \mathscr{W}_\nu ] .
\label{F-decomposition}
\end{align}
By substituting the decomposition (\ref{F-decomposition}) into the $SU(2)$ Yang-Mills-Higgs Lagrangian, we obtain 
\begin{align}
   \mathscr{L}_{\rm YMH} 
=& -\frac{1}{4} \mathscr{F}_{\mu\nu}[\mathscr{V}] \cdot \mathscr{F}^{\mu\nu}[\mathscr{V}]
\nonumber\\&
- \frac{1}{4} (\mathscr{D}_\mu[\mathscr{V}] \mathscr{W}_\nu - \mathscr{D}_\nu[\mathscr{V}] \mathscr{W}_\mu)^2
+ \frac{1}{2 } M_{W}^2 \mathscr{W}^\mu  \cdot \mathscr{W}_\mu 
\nonumber\\&
+ \frac{1}{2}  \mathscr{F}_{\mu\nu}[\mathscr{V}] \cdot ig[ \mathscr{W}^\mu , \mathscr{W}^\nu ]
- \frac14 (i g [ \mathscr{W}_\mu , \mathscr{W}_\nu ])^2
 ,
\label{L-VW}
\end{align}
where each term is $SU(2)$ invariant. 
Then it is easy to observe that the vector field $\mathscr{W}_\mu$ has the ordinary kinetic term and the mass term.  Therefore, there is a massive vector pole in the propagator of $\mathscr{W}_\mu$ (after a certain gauge fixing). Thus, $\mathscr{W}_\mu$ is not an auxiliary field, but is a propagating field with the mass $M_{W}$  (up to possible quantum corrections).


\section{Confined massive phase: SU(2) case}

Remarkably,  the  field strength $\mathscr{F}_{\mu\nu}[\mathscr{V}](x)
 :=  \partial_\mu \mathscr{V}_\nu(x) - \partial_\nu \mathscr{V}_\mu(x) -ig [\mathscr{V}_\mu(x) , \mathscr{V}_\nu(x) ]$ of   $\mathscr{V}_\mu(x)$ 
is  shown to be proportional to $\hat{\bm{\phi}}(x)$ 
\cite{KKSS15}:
\begin{align}
 \mathscr{F}_{\mu\nu}[\mathscr{V}](x)
 =&  \hat{\bm{\phi}}(x)  \{ \partial_\mu c_\nu(x) - \partial_\nu c_\mu(x) + H_{\mu\nu}(x) \} ,
\nonumber\\ 
 H_{\mu\nu}(x) :=& ig^{-1} \hat{\bm{\phi}}(x)  \cdot [\partial_\mu \hat{\bm{\phi}}(x)  , \partial_\nu \hat{\bm{\phi}}(x)   ] .
 \label{C26-F-decomp1}
\end{align}
We can introduce  the Abelian-like $SU(2)$ gauge-invariant field strength $f_{\mu\nu}(x)$ by 
\begin{align}
  f_{\mu\nu}(x)  :=&  \hat{\bm{\phi}}(x) \cdot \mathscr{F}_{\mu\nu}[\mathscr{V}](x) 
\nonumber\\ 
=& \partial_\mu c_\nu(x) - \partial_\nu c_\mu(x) + H_{\mu\nu}(x) .
  \label{f-def-SU2}
\end{align}
In the low-energy $E \ll M_{W}$ or the long-distance $r \gg M_{W}^{-1}$ region, we can neglect the field $\mathscr{W}_\mu$ as the first approximation.
Then the dominant low-energy modes are described by the restricted Lagrangian density:
\begin{align}
   \mathscr{L}_{\rm YM}^{\rm rest}
=  - \frac{1}{4} \mathscr{F}^{\mu\nu}[\mathscr{V}] \cdot \mathscr{F}_{\mu\nu}[\mathscr{V}]
= - \frac{1}{4} f^{\mu\nu} f_{\mu\nu} .
\label{restricted-YM}
\end{align}
The resulting gauge theory with the Lagrangian (\ref{restricted-YM}) is called the \textit{restricted Yang-Mills theory}.
Consequently, the $SU(2)$ Yang-Mills theory looks like the Abelian gauge theory (\ref{restricted-YM}). 
But, even at this stage the original non-Abelian gauge symmetry $SU(2)$ is not broken. 


In the low-energy $E \ll M_{W}$ or the long-distance $r \gg M_{W}^{-1}$ region, the massive components $\mathscr{W}_\mu(x)$ become negligible and the restricted theory  become dominant. 
This is equal to a phenomenon called the \textit{``Abelian'' dominance} \cite{tHooft81,EI82} in quark confinement.
We have shown that the ``Abelian'' dominance in quark confinement of the Yang-Mills theory is understood as a consequence of the Higgs mechanism defined in a gauge-invariant way for the relevant (or equivalent) Yang-Mills-Higgs model.
The Abelian dominance was confirmed  for the string tension \cite{SY90} and for the propagator \cite{AS99,BCGMP03} for the $SU(2)$ Yang-Mills theory on the lattice in the Maximal Abelian gauge \cite{KLSW87}, and later reconfirmed based on the gauge-invariant formulation on the lattice  for the string tension \cite{KKS14} and  the full propagator \cite{SKKMSI07}.

Notice that $H_{\mu\nu}(x)$ is {\it locally} closed $(dH=0)$ and hence it can be {\it locally} exact $(H=dh)$  due to the Poincare$\acute{}$ lemma. Then $H_{\mu\nu}(x)$ has the Abelian  potential $h_\mu(x)$: 
\begin{align}
H_{\mu\nu}(x) 
 =&  \partial_\mu h_\nu(x) - \partial_\nu h_\mu(x) 
 .
 \label{C26-magnetic-potential}
\end{align}
Therefore, the  $SU(2)$ gauge-invariant Abelian-like field strength 
$f_{\mu\nu}$ is rewritten as 
\begin{align}
  f_{\mu\nu}(x)  
=& \partial_\mu G_\nu(x)  - \partial_\nu G_\mu(x)  ,
\  
G_\mu(x) :=c_\mu(x)+h_\mu(x) .
\end{align}
We call $c_\mu$ the \textit{electric potential} and $h_\mu$ the \textit{magnetic potential}. 
Indeed, $h_\mu$ agrees with the Dirac magnetic potential, see section 6.10 of  \cite{KKSS15}. 

We can define the \textit{magnetic--monopole current} $k^\mu(x)$   in a gauge-invariant way:
\begin{equation}
 k^\mu(x) = \partial_\nu  {}^{\displaystyle *}f^{\mu\nu}(x) , 
\end{equation}
where $ {}^{\displaystyle *}$ denotes the Hodge dual, e.g., for $D=4$, the dual tensor ${}^{\displaystyle *}f^{\mu\nu}$ of $f^{\mu\nu}$ is defined by
$
  {}^{\displaystyle *}f^{\mu\nu}(x)  :=  \frac12 \epsilon^{\mu\nu\rho\sigma} f_{\rho\sigma}(x) .
$
The magnetic current $k^\mu(x)$ is not identically zero, since the Bianchi identity valid for the electric potential $c_\mu$ is violated by the magnetic potential $h_\mu$. 
The contribution of the gauge-invariant magnetic monopole to the Wilson loop average can be detected using the non-Abelian Stokes theorem for the Wilson loop operator, see \cite{Kondo08,MK15} and section 6 of \cite{KKSS15}. 

The restricted Yang-Mills theory obtained from the original $SU(2)$ Yang-Mills theory 
has the magnetic part besides the electric part which exists in  the usual non-compact $U(1)$ gauge theory.
Therefore, the restricted Yang-Mills theory is reagarded as the continuum counterpart to the compact $U(1)$ gauge theory on the lattice   which involves the magnetic monopoles leading to confinement in the strong coupling region   \cite{BMK77,Polyakov75}.
It is known \cite{BKSLN82,CJR83} that the compact $U(1)$ gauge theory on the lattice has two phases: confinement phase due to magnetic monopoles in the strong coupling region \cite{BMK77,Polyakov75} which is separated  by a critical coupling from the Coulomb phase in the weak coupling region \cite{FS82,Guth80}.

The Yang-Mills-Higgs model includes the parameters specifying the potential besides the gauge coupling.  They are  arbitrary and hence the mass gap of the theory is not uniquely determined. 
In sharp contrast to the Yang-Mills-Higgs model,  the mass gap in the Yang-Mills theory should be generated in a dynamical way without breaking gauge invariance, and it is determined without free parameters to be adjusted. 

In the Yang-Mills theory, indeed, the mass $M_{W}$ can be generated in a dynamical way, e.g., by a gauge-invariant vacuum condensation $\langle \mathscr{W}^\mu  \cdot \mathscr{W}_\mu \rangle$ so that  $M_{W}^2 \simeq \langle \mathscr{W}^\mu  \cdot \mathscr{W}_\mu  \rangle$ due to the quartic self-interactions  
$-\frac14 ( ig[\mathscr{W}_\mu(x),\mathscr{W}_\nu(x)])^2$ 
among $\mathscr{W}_\mu(x)$ field, in sharp contrast to the ordinary Yang-Mills-Higgs model.  
The analytical calculation for such a condensate was done in \cite{Kondo06}. 
Moreover,  the mass $M_{W}$ has been measured by 
  numerical simulations  on the lattice in \cite{SKKMSI07} (see also section 9.4 of \cite{KKSS15}) as 
\begin{align}
  M_{W} &\simeq 2.69  \sqrt{\sigma_{\rm phys}} \simeq  
1.19 
 {\rm GeV} ,
\end{align}
where $\sigma_{\rm phys}$ is the string tension of the linear potential in the quark-antiquark potential.

The mass $M_W$ is used to show the existence of  confinement-deconfinement phase transition at a finite critical temperature $T_c$, separating confinement phase with vanishing Polyakov loop average at low temperature and deconfinement phase with non-vanishing Polyakov loop average at high temperature \cite{Kondo15}. The critical temperature $T_c$ is obtained from the calculated ratio $T_c/M_W$ for a given $M_W$, which provides a reasonable estimate.

Notice that we cannot introduce the ordinary mass term for the  field $\mathscr{V}_\mu$, since it breaks the original gauge invariance. 
But, another mechanism of generating mass for the Abelian gauge field $G_\mu :=c_\mu+h_\mu$ could be available, e.g.,  magnetic mass for photon due to the Debye screening caused by magnetic monopoles, which yields confinement and mass gap in three-dimensional Yang-Mills-Higgs theory as shown in \cite{Polyakov77b}.
Moreover, the Abelian gauge field must be confined, which is a problem of gluon confinement. 
In view of these, the full propagator of the Abelian gauge field  must have a quite complicated form, as has been discussed in e.g., \cite{Kondo11}.

In the Yang-Mills-Higgs model, the gauge field $\mathscr{A}_\mu$ and the scalar field $\bm{\phi}$ are independent field variables. 
However, the Yang-Mills theory should be described by the gauge field $\mathscr{A}_\mu$ alone and hence the scalar field $\bm{\phi}$ must be supplied by the gauge field $\mathscr{A}_\mu$  due to the strong interactions. In other words, the scalar field $\bm{\phi}$ should be given as a (complicated) functional of the gauge field. This is achieved by imposing the constraint which we call the  \textit{reduction condition} \cite{KMS06,KMS05}, see also section 4 of \cite{KKSS15}.  
We choose  e.g., 
\begin{align}
  \bm{\chi}(x) := [\hat{\bm{\phi}}(x) , \mathscr{D}^{\mu}[\mathscr{A}]  \mathscr{D}_{\mu}[\mathscr{A}] \hat{\bm{\phi}}(x) ] = \bm{0}  
,
  \label{reduction-mYM2}
\end{align}
which is also written as $\mathscr{D}^{\mu}[\mathscr{V}] \mathscr{W}_\mu(x) = 0$.
This condition is gauge covariant, 
\begin{align}
  \bm{\chi}(x) \to U(x) \bm{\chi}(x) U^{-1}(x)
  .
\end{align}
This is easily shown from the gauge transformation (\ref{gauge-transf}) of the scalar field and the Yang-Mills field. 

The reduction condition plays the role of eliminating the extra degrees of freedom introduced by the radially fixed scalar field into  the Yang-Mills theory \cite{KKSS15}. 
The reduction condition represents as many conditions  as the independent degrees of freedom of the radially fixed scalar field $\bm{\phi}(x)$, since 
\begin{align}
  \bm{\chi}(x) \cdot \hat{\bm{\phi}}(x) = 0 .
\end{align}
Therefore, imposing the  reduction condition (\ref{reduction-mYM2}) exactly eliminates extra degrees of freedom introduced by the radially fixed scalar field (\ref{SU2-YMH-1-constraint}), see \cite{KKSS15}.

Fortunately, the reduction condition is automatically satisfied in the level of  field equations. 
We introduce a Lagrange multiplier field $\lambda(x)$ to incorporate the constraint (\ref{SU2-YMH-1-constraint}) into the Lagrangian:
\begin{align}
\mathscr{L}_{\rm YMH}^\prime  
=  
\mathscr{L}_{\rm YMH} + \lambda(x) \left( \bm{\phi}(x) \cdot \bm{\phi}(x) - v^2 \right)
 .
\label{mYM3}
\end{align}
Then the field equations are obtained as
\begin{align}
\frac{\delta S_{\rm YMH}^\prime}{\delta \lambda(x)} 
=& \bm{\phi}(x) \cdot \bm{\phi}(x) - v^2 = 0 , 
\label{field-eq-tHP1b}
\\
\frac{\delta S_{\rm YMH}^\prime}{\delta \mathscr{A}^{\mu }(x)} 
=& \mathscr{D}^{\nu}[\mathscr{A}] \mathscr{F}_{\nu\mu} (x) -  ig [ \bm{\phi} (x) ,  \mathscr{D}_{\mu}[\mathscr{A}] \bm{\phi}(x) ]  = 0 ,
\label{field-eq-tHP1a}
\\
\frac{\delta S_{\rm YMH}^\prime}{\delta \bm{\phi}(x)} 
=& -  \mathscr{D}^{\mu}[\mathscr{A}] \mathscr{D}_{\mu}[\mathscr{A}] \bm{\phi}(x) 
-  2 \bm{\phi}(x)V^\prime(\bm{\phi}(x) \cdot \bm{\phi}(x))
\nonumber\\&
 +  2\lambda(x) \bm{\phi}(x) 
= 0 .
\label{field-eq-tHP1c}
\end{align}
The reduction condition (\ref{reduction-mYM2}) follows by applying the covariant derivative $\mathscr{D}^{\mu}[\mathscr{A}] $ to (\ref{field-eq-tHP1a}) as 
$\mathscr{D}^{\mu}[\mathscr{A}] \mathscr{D}^{\nu}[\mathscr{A}] \mathscr{F}_{\nu\mu}  =ig \mathscr{D}^{\mu}[\mathscr{A}]  [ \bm{\phi}  ,  \mathscr{D}_{\mu}[\mathscr{A}] \bm{\phi}  ]
= ig [ \bm{\phi}  ,  \mathscr{D}^{\mu}[\mathscr{A}] \mathscr{D}_{\mu}[\mathscr{A}] \bm{\phi} ]$ (this is the covariant version of the current conservation law), since 
$\mathscr{D}^{\mu}[\mathscr{A}] \mathscr{D}^{\nu}[\mathscr{A}] \mathscr{F}_{\nu\mu}  =0$. 
Taking the commutator  of the field equation (\ref{field-eq-tHP1c}) for the scalar field $\bm{\phi}$  with $\bm{\phi}$, we find that the reduction condition (\ref{reduction-mYM2}) is automatically satisfied, irrespective of the choice of the potential function   $V(\bm{\phi} \cdot \bm{\phi} )$: 
 $[ \bm{\phi} , \mathscr{D}^{\mu}[\mathscr{A}] \mathscr{D}_{\mu}[\mathscr{A}] \bm{\phi}]
=[ \bm{\phi},-  2 \bm{\phi} V^\prime(\bm{\phi} \cdot \bm{\phi} ) +  2\lambda  \bm{\phi} ]=0$.

Notice that the equivalence between the Yang-Mills-Higgs theory and the pure Yang-Mills theory is expected to hold only when the scalar field is radially fixed. 
If we include the radial degree of freedom for the scalar field, the equivalence is lost. 
Indeed, the radial degree of freedom for the scalar field corresponds to the Higgs particle with a non-zero mass. 

\section{Conclusion and discussion}

In this paper we have given a gauge-independent description for the Higgs mechanism by which a gauge boson acquires the mass in a manifestly gauge-invariant way. 
We have written the resulting massive gauge modes $\mathscr{W}_\mu$ explicitly in the operator level. 
Therefore, we can describe the Higgs mechanism without assuming spontaneous breakdown of gauge symmetry relying on a non-vanishing vacuum expectation value of the scalar field. 
In this way, we can understood the mass generation of gauge bosons in the gauge-invariant way without breaking the original gauge symmetry. 
The spontaneous symmetry breaking is sufficient but not necessary for the Higgs mechanism to work.

The novel description of the Higgs mechanism enables us to discuss the  confinement-Higgs complementarity from a new perspective.
Our results suggest that 
the $SU(2)$ Yang-Mills theory in the gapped or massive phase  is equivalent to the Yang-Mills-Higgs theory with a radially fixed adjoint scalar field in the Higgs phase which is conventionally considered to be associated to the spontaneous symmetry breaking $G=SU(2) \to  {H}=U(1)$. 
The gapped or massive phase is regarded as the confinement phase, which was confirmed on a lattice by numerical simulations for the reformulated Yang-Mills theory \cite{KKSS15}.

Moreover, we have discussed the implications of the gauge-invariant Higgs  mechanism for quark confinement. 
We have shown that the ``Abelian'' dominance  in quark confinement of the $SU(2)$ Yang-Mills theory is understood as a consequence of the gauge-invariant Higgs phenomenon  for the relevant $SU(2)$ Yang-Mills-Higgs model.

The case of larger gauge groups  $SU(N)$ ($N \ge 3$)   will be treated in a subsequent paper. 
In particular, some interesting cases $SU(3) \to U(1) \times U(1)$, $SU(3) \to U(2)$, and $SU(2) \times U(1) \to U(1)$ will be discussed in detail. 



\begin{acknowledgements}

This work is  supported by Grants-in-Aid for Scientific Research (C) No.15K05042 from the Japan Society for the Promotion of Science (JSPS).

\end{acknowledgements}

\appendix
\section{Higgs mechanism for the complete SSB}

We consider the {Abelian-Higgs theory} or  {$U(1)$ gauge-scalar theory} with the Lagrangian density:
\begin{align}
\mathscr{L}_{\rm AH} =& - \frac{1}{4} F_{\mu\nu} F^{\mu\nu} +  ( D_{\mu}\phi)^{*} (D^{\mu}\phi) - V(\phi^{*} \phi) ,  
\nonumber\\
 V(\phi^{*} \phi) =&  \frac{\lambda}{2} \left( \phi^{*} \phi - \frac{\mu^{2}}{\lambda}  \right)^{2} , \ \phi \in \mathbb{C} ,  
\ \lambda>0 ,
\end{align}
where 
$F_{\mu\nu}(x)=\partial_\mu A_\nu(x) - \partial_\nu A_\mu(x)$ is the  field strength of the $U(1)$ gauge field $A_\mu(x)$  and $D_{\mu}=\partial_{\mu}-ieA_{\mu}(x)$ is a $U(1)$ covariant derivative for the complex scalar field $\phi(x) \in \mathbb{C}$ with $q$ being the electric charge of $\phi(x)$. 
Here $*$ denotes the complex conjugate.
For $\mu^2>0$, the minimum of the potential is attained when the magnitude of the scalar field is equal to the value: 
\begin{equation}
		|\phi(x)| = \frac{v}{\sqrt{2}}  , \quad v = \sqrt{\frac{\mu^2}{\lambda/2}} .
\end{equation}

If we use a representation of  {polar decomposition} for the  radially fixed scalar field: 
\begin{align}
	&	\phi(x) = \frac{v}{\sqrt{2}} e^{i \pi(x)/v} \in \mathbb{C} , \quad \pi(x) \in \mathbb{R} ,
	\label{polar}
\end{align}
the covariant derivative reads 
\begin{align}
D_\mu \phi=(\partial_{\mu} - ie A_{\mu})\phi(x) 
= - \frac{v}{\sqrt{2}} ie  \left( A_{\mu} - \frac{1}{ev} \partial_{\mu} \pi    \right) e^{i \pi/v} ,
\end{align}
and the kinetic term of the scalar field reads
\begin{align}
		(D_{\mu} \phi)^{*} (D^{\mu} \phi) 
=&   \frac12 e^2 v ^2 \left( A_{\mu} - \frac{1}{ev} \partial_{\mu} \pi   \right)^2 .
\end{align}

By introducing a new (massive) vector field $W_{\mu}$ by
\begin{equation}
		W_{\mu}(x) := A_{\mu}(x) - m^{-1} \partial_{\mu} \pi(x) , \ m := ev, 
		\label{Proca}
\end{equation}
 $\mathscr{L}_{\rm AH}$ is completely rewritten in terms of   $W_{\mu}$:
\begin{align}
 \mathscr{L}_{\rm{AH}} = -\frac{1}{4} (\partial_{\mu} W_{\nu} - \partial_{\nu} W_{\mu})^2 + \frac{1}{2} m^2 W_{\mu} W^{\mu} . 
\end{align}
The field $\pi$ is usually interpreted as the massless Nambu-Goldstone boson  associated with the complete SSB $G=U(1) \to H=\{ 1 \}$, which is absorbed into the massive field $W_{\mu}$.  
For $G=U(1)$, we find that the massive vector field $W_{\mu}$ has a manifestly gauge-invariant representation written in terms of $A_\mu$ and $\phi$   ($\hat{\phi}:=\phi(x)/|\phi(x)|$): 
\begin{equation}
 W_{\mu}(x)   
= ie^{-1} \hat{\phi}^*(x)  D_\mu \hat{\phi}(x) 
= -ie^{-1}  \hat{\phi}(x)  D_\mu \hat{\phi}^*(x)   .
\label{U1}
\end{equation}
This reduces to (\ref{Proca}) for the parameterization (\ref{polar}). 
The representation (\ref{U1}) is \textit{independent from the parameterization} of the scalar field. 
Therefore, a different representation is obtained from another parameterization:
\begin{align}
	\phi(x) = \frac{1}{\sqrt{2}} [v+\varphi(x) + i\chi(x)] .
\end{align}


\end{document}